\documentclass[draftcls,12pt,onecolumn,dvipdfm]{IEEEtran}
\usepackage{mathrsfs}
\usepackage{amsmath}
\usepackage{amssymb}
\usepackage{enumerate}
\usepackage{color,xcolor}
\usepackage{graphicx}
\usepackage{subfigure}
\usepackage{pifont}
\usepackage{algorithmic}

\ifCLASSINFOpdf

\else

\fi

\newtheorem{definition}{Definition}

\newtheorem{algorithm}{\textbf{Algorithm}}

\usepackage[square,numbers,sort&compress]{natbib}

\usepackage{ifpdf}

\begin{document}

\title{Bounds on the ML Decoding Error Probability of RS-Coded Modulation over AWGN Channels}

\author{Qiutao~Zhuang,
        Xiao~Ma,~\IEEEmembership{Member,~IEEE},
        and~Aleksander~Kav\v{c}i\'{c},~\IEEEmembership{Member,~IEEE}
\thanks{Q.~Zhuang and X.~Ma are with the Department of Electronics and Communication
Engineering, Sun Yat-sen University, Guangzhou 510275, China. (Email:
maxiao@mail.sysu.edu.cn)}
\thanks{A.~Kav\v{c}i\'{c} is with the Department of Electrical Engineering, University of Hawaii at Manoa, Honolulu, 96822 HI USA.}
\thanks{This work was supported by the 973 Program (No. 2012CB316100), by the NSF (No. 61172082) of China and by NSF grant CCF-1018984. When this work was conducted, X.~Ma and A.~Kav\v{c}i\'{c} were visiting scholars supported by the Institute of Networking Coding at Chinese University of Hong Kong.}
}


\maketitle

\begin{abstract}

This paper is concerned with bounds on the maximum-likelihood~(ML) decoding error probability
of Reed-Solomon~(RS) codes over additive white Gaussian noise~(AWGN) channels. To resolve the
difficulty caused by the dependence of the Euclidean distance spectrum on the way of signal mapping, we propose to use random mapping, resulting in an ensemble of RS-coded modulation~(RS-CM) systems. For this ensemble of RS-CM systems, analytic bounds are derived, which can be evaluated from the known~(symbol-level) Hamming distance spectrum.  Also
presented in this paper are simulation-based bounds, which are applicable to any specific RS-CM system and can be evaluated by the aid of a list decoding~(in the Euclidean space) algorithm. The simulation-based bounds do not need distance spectrum and are numerically tight for short RS codes in the regime where the word error rate~(WER) is not too low. Numerical comparison results are relevant in at least three aspects. First, in the short code length
regime, RS-CM using BPSK modulation with random mapping has a better performance than binary random linear codes. Second, RS-CM with random mapping~(time varying) can have a better performance than with specific mapping. Third, numerical results show that the recently proposed Chase-type decoding algorithm is essentially the ML decoding algorithm for short RS codes.

\end{abstract}


%
\IEEEpeerreviewmaketitle

\section{Introduction}\label{introduction}
%
%
%
%

\IEEEPARstart{R}eed-Solomon~(RS) codes are an important class of
algebraic codes, which have been widely used in many practical
systems, including space and satellite communications, data storage,
digital audio/video transmission and file
transfer~\cite{Costello98}. The widespread use of RS codes is
primarily due to their excellent error-correction capability, a
consequence of their maximum distance separable~(MDS) property.
Hence investigating the decoding algorithms for RS codes is important in
both practice and theory. The traditional hard-decision decoding~(HDD) algorithms, say the
Berlekamp-Massey~(BM) algorithm~\cite{Berlekamp68}, are
efficient to find the unique codeword~(if it exists) within a Hamming sphere of radius less than the
half minimum Hamming distance. Hence, their error-correction capability is limited by the half
minimum Hamming distance bound. In contrast, Guruswami-Sudan~(GS) algorithm~\cite{Sudan97}\cite{Guruswami99} can
enlarge the decoding radius and may output a list of candidate codewords. Hence, GS algorithm
can correct errors beyond the half minimum Hamming distance bound. To further improve the
performance, one needs turn to the soft-decision decoding~(SDD) algorithms.


The SDD algorithms with feasible complexity for RS codes include the generalized minimum distance~(GMD) algorithm~\cite{Forney66b}, the Chase-GMD algorithm~\cite{Tang01}, the Koetter-Vardy~(KV) algorithm~\cite{Koetter03}, the Chase-KV algorithm~\cite{Zhang12}, the ordered statistic decoding~(OSD) algorithm~\cite{Fossorier95}\cite{Jin08}, and the adaptive belief propagation~(ABP) algorithm~\cite{Jiang06}, etc. Recently, two Chase-type decoding algorithms have been proposed for RS codes~\cite{Bellorado10}\cite{Tang13x}. All these efforts have been made to improve incrementally the performance of some existed algorithms and to achieve the performance of the maximum likelihood decoding though it has been shown by Guruswami and Vardy in~\cite{Guruswami05} that maximum likelihood~(ML) decoding of a general RS code is NP-hard\footnote{For short RS codes with binary phase-shift keying~(BPSK) signalling, ML decoding can be performed based on the algebraic structure of their binary images~\cite{Vardy91}\cite{Ponnampalam02}.}.

An immediate question is how to measure the sub-optimality~(the gap to the ML decoding) of varieties of decoding algorithms. Though the ML decoding algorithm is
prohibitively complex, tight bounds can be used to predict the
performance without resorting to computer simulations. As mentioned
in~\cite{Sason06}, most bounding techniques have connections to
either the 1965 Gallager
bound~\cite{Duman98,Duman98a,Twitto07} or the 1961
Gallager
bound~\cite{Sphere94,TSB94,Divsalar99,Divsalar03,Yousefi04,Yousefi04a,Mehrabian06,Ma13,Zhuang13x}
based on Gallager's first bounding technique~(GFBT). For a RS-coded modulation~(RS-CM) system even with BPSK signalling, these bounds become helpless since there is no simple way to derive the bit-level Hamming weight spectrum from the known symbol-level Hamming weight distribution. The difficulty is partially caused by the dependence of the bit representation of a symbol on the choice of basis. In~\cite{Ponnampalam02}, upper bounds on ML decoding error
probability were computed by computer search of the bit-level weight spectrum for short RS codes, while in~\cite{Mostafa04}, upper bounds were derived from the average binary weight enumerator~(BWE) of the RS codes. Generally, for RS-CM using
other modulations~\cite{Yar08}, the
Euclidean distance spectrum of the codewords depends on the chosen signal mapping,
resulting in the difficulty of performance evaluation.

At this point, we emphasize that a study of a communication system that utilizes RS codes is incomplete if only the coding aspect is considered. The performance of any communication system heavily depends on the chosen modulation and constellation mapping. It is well known that, even if the signalling constellation is fixed, the system performance still depends on which exact mapping from coded symbols to constellation points is chosen. For this reason, in this paper, our goal is to consider the analysis of the ML decoding performance of a RS code in conjunction with a high order modulation. However, incorporating a constellation mapping structure into the analysis of ML decoding of RS codes seems to only further complicate the analysis. This could be one reason why most bounds were developed for binary codes with BPSK modulation.

To resolve this difficulty, instead of considering any specific modulation mapping, we adopt a {\em random} mapping approach which gives rise to an {\em ensemble} of RS-CM systems. For such an {\em ensemble}, we will show that it is indeed possible to find {\em analytic} bounds on the performance of ML decoding error probability. {\em Randomization} is a powerful technique to analyze the performance, which has been widely used in the field of informaiton and coding theory. For example, Benedetto~\cite{Benedetto96} introduced {\em random interleaver} to derive the weight distribution of an ensemble of turbo codes, while Richardson-Urbanke~\cite{Richardson01} and Luby~\cite{Luby98b} {\it et al} showed how to predict the performance of LDPC codes by introducing the random irregular Tanner graphs. However, in our approach, it is not the code that is random (indeed, the code is a well-constructed algebraic code), but it is the modulation mapping (which maps symbols to constellation points) that is random. For this ensemble of RS-CM systems, analytic bounds are derived based on the results in~\cite{Zhuang13x}, which require only the known~(symbol-level) Hamming distance spectrum of the RS code given that the signal constellation is fixed. To the best of our knowledge, a randomized approach to modulation mapping has never been used in the past as an analytic tool to study the performance of modulation mappings and this presents the first contribution of this paper.

The second contribution of this paper is that we present simulation-based bounds which are applicable to any specific RS-CM system. The simulation bounds can be evaluated by the aid of a list decoding~(in the Euclidean space) algorithm. For short codes, the bounds are tight (almost overlapped). This also shows that the recently proposed tree-based Chase-type algorithm~\cite{Tang13x} is near optimal for short RS codes.

The rest of this paper is organized as follows. In Sec.~\ref{sec2}, the union bound~(UB) and the sphere bound~(SB) for general codes are reviewed. In Sec.~\ref{sec3}, we propose analytic bounds for the
ensemble of RS-CM systems using random modulation and derive the
corresponding average Euclidean distance enumerating function. In Sec.~\ref{sec4}, we present
simulation-based bounds for specific RS-CM system using the aid of a list decoding algorithm. Numerical results are presented in
Sec.~\ref{sec5}, and Sec.~\ref{conclusion} concludes this paper.

\section{RS-coded Modulation}\label{sec2}

\subsection{System Model}

Let $\mathbb{F}_q\stackrel{\Delta}{=}\{\alpha_0,\alpha_1,\cdots,\alpha_{q-1}\}$
be the finite field of size $q$. A codeword of an RS code
$\mathcal{C}_q[n, k,d_{\min}]$ with length $n$, dimension $k$ and minimum Hamming distance
$d_{\min}=n-k+1$ can be obtained by evaluating a polynomial of degree
less than $k$ over a set of $n$ distinct points, denoted by
$\mathcal{P}\stackrel{\Delta}{=}\{\beta_0,\beta_1,\cdots,\beta_{n-1}\}\subseteq
\mathbb{F}_q$.

{\em Encoding:} Let $\underline{u}=(u_0,u_1,\cdots,u_{k-1})\in
\mathbb{F}_q^k$ be an information sequence to be transmitted, which specifies a message polynomial $u(x) = u_0 + u_1x + \cdots + u_{k-1}x^{k-1}$. The corresponding codeword is then given by
\begin{equation}
\underline{c}=(c_0, c_1, \cdots,
c_{n-1})=(u(\beta_0),u(\beta_1),\cdots,u(\beta_{n-1})).
\end{equation}

{\em Mapping:} The codeword $\underline{c}$ is transformed into a
signal vector $\underline s=(s_0, s_1, \cdots, s_{n-1})$, where
$s_i=\phi(c_i)\in \mathbb{R}^\ell$ is an $\ell$-dimensional signal
which is determined by the mapping rule $\phi$. The constellation
$\mathcal {X}\stackrel{\Delta}{=}\{\phi(\alpha),\alpha
\in\mathbb{F}_q\}$ is of size $q$, whose form depends on the
modulation scheme. For an example, we consider a $64$-ary RS code. If BPSK is implemented, we have $\ell=6$ and $\mathcal {X}=\{-1,+1\}^6$, the six-fold Cartesian product of $\{-1, +1\}$; if $8$-PSK is
implemented, we have $\ell=4$ and $\mathcal {X}=\{8$-PSK$\}^2$; while if $64$-QAM is implemented, we have $\ell=2$ and $\mathcal {X}=\{\pm 1, \pm 3,\pm 5,\pm 7\}^2$. All signal vectors are collectively denoted by $\mathcal{S} \stackrel{\Delta}{=} \{\underline s\mid s_t = \phi(c_t), 0\leq t \leq n-1, {\underline c}\in \mathcal{C}_q\}$.

{\em Channel:} Assume that the signal vector $\underline s$ is
transmitted through an AWGN channel. The received vector is denoted
by $\underline{y} = {\underline s} + {\underline z}$, where
$\underline z$ is a sample from a white Gaussian noise process with
zero mean and double-sided pow spectral density $\sigma^2$.

{\em ML Decoding:} Assume that each codeword is transmitted with equal probability. The optimal decoding that minimizes the word-error probability is the ML
decoding, which, for AWGN channels, is equivalent to finding the nearest signal vector
$\hat{\underline s} \in \mathcal{S}$ to $\underline y$.

Hereafter, we may not distinguish $\underline c$ from $\underline s$ when
representing a codeword of RS-CM.

\subsection{Distance Enumerating Functions of RS-CM}

The weight enumerating function of a RS code $\mathcal{C}_q[n,
k,d_{\min}]$ is defined as
\begin{equation}
    W(X) \stackrel{\Delta}{=} \sum_{i=d_{\min}}^{n}W_{i}X^{i},
\end{equation}
where $X$ is a dummy variable and $W_i$ denotes the number of
codewords having Hamming weight $i$,  which can be determined by~\cite{Moon05}
\begin{equation}
W_i=\binom{n}{i}(q-1)\sum_{j=0}^{i-d_{\min}}(-1)^{j}\binom{i-1}{j}q^{i-j-d_{\min}},i\geq
d_{\min}.
\end{equation}

Given a codeword $\underline s$ of RS-CM, we denote $A_{\delta|\underline s}$ the number of codewords having the Euclidean distance
$\delta$ with $\underline s$. We define
\begin{equation*}
  A_{\delta} = \frac{1}{q^k}\sum_{\underline s} A_{\delta|\underline s},
\end{equation*}
which is the average number of {\em ordered} pairs of codewords with Euclidean distance $\delta$.

\begin{definition}
The {\em Euclidean distance enumerating function} of RS-CM is
defined as~\cite[(2)]{Zhuang13x}
\begin{equation}\label{A(X)}
A(X) \stackrel{\Delta}{=} \sum_{\delta}A_{\delta}X^{\delta^2},
\end{equation}
where $X$ is a dummy variable and the summation is over all possible
distance $\delta$. We call $\{A_{\delta}\}$ the Euclidean distance
spectrum.
\end{definition}

\subsection{Upper Bounds for RS-CM}
The conventional union bound~(UB) on the ML decoding error
probability ${\rm Pr}\{E\}$ for RS-CM is
\begin{equation}\label{UB}
{\rm Pr} \{E\}\leq
\sum_{\delta}A_{\delta}Q\left(\frac{\delta}{2\sigma}\right),
\end{equation}
where $Q\left(\frac{\delta}{2\sigma}\right)$ is the pair-wise error
probability with
\begin{equation}
Q(x)\stackrel{\Delta}{=}\int
_{x}^{+\infty}\frac{1}{\sqrt{2\pi}}e^{-\frac{z^{2}}{2}}\,{\rm d}z.
\end{equation}

The UB can be tightened by the use of the following sphere bound~(SB) as shown in~\cite[(26)]{Zhuang13x},
\begin{equation}\label{SB}
{\rm Pr} \{E\}\leq
\int_{0}^{+\infty}\min\left\{f_u(r),1\right\}g(r)~{\rm d}r,
\end{equation}
where
\begin{equation}
f_u(r)=\sum_{\delta}A_{\delta} p_2(r,
    \delta),
\end{equation}
\begin{equation}\label{p2}
    p_2(r, \delta) = \left\{\begin{array}{rl}
                                         \frac{\Gamma(\frac{\ell n}{2})}{\sqrt{\pi}~\Gamma(\frac{\ell n-1}{2})}\int_{0}^{\arccos(\frac{\delta}{2r})}\sin^{\ell n-2}
\phi~{\rm d}\phi, & r > \frac{\delta}{2} \\
                                           0, & r \leq \frac{\delta}{2}
                                         \end{array}\right.,
\end{equation}
and
\begin{equation}\label{gr}
g(r)=\frac{2r^{\ell
n-1}e^{-\frac{r^{2}}{2\sigma^{2}}}}{2^{\frac{\ell n}{2}}\sigma^{\ell
n}\Gamma(\frac{\ell n}{2})},~~r\geq 0,
\end{equation}
which is determined by the Euclidean distance spectrum
$\{A_{\delta}\}$.

\section{Analytic Bounds for an Ensemble of RS-coded Modulation Systems}\label{sec3}

As seen from Sec.~\ref{sec2}, computing the derived upper bounds on the ML decoding error probability for RS-CM requires the Euclidean distance spectrum $\{A_{\delta}\}$, which depends on the way of signal mapping $\phi$ and is usually difficult to compute. To resolve this difficulty, in this section, we propose to use random mapping, resulting in an ensemble of RS-CM systems. For this ensemble of RS-CM systems, analytic bounds are derived, which can be evaluated from the weight enumerating function $W(X)$.

\subsection{Average Euclidean Distance Enumerating Function of RS-CM with Random Modulation}

Let $\mathcal{C}_q[n, k, d_{\min}]$ be an RS code defined over $\mathbb{F}_q$ and $\mathcal{X} \subset \mathbb{R}^{\ell}$ be a signal constellation of size $q$. Let $\Phi=\{\phi^{(1)},\phi^{(2)},\cdots,\phi^{(q!)}\}$ be the set of all one-to-one mapping rules from $\mathbb{F}_q$ to $\mathcal{X}$. Assume that ${\underline \phi} = (\phi_0, \phi_1, \cdots, \phi_{n-1})$ is a random sequence, whose components are sampled independently and uniformly from $\Phi$.  Define
$\mathcal{S}(\mathcal{C}_q,{\underline \phi})\stackrel{\Delta}{=}\left\{
\underline{s}\mid s_t=\phi_t(c_t), 0\leq
t\leq n-1, \underline{c}\in \mathcal{C}_q \right\}$. It can be seen that $\mathcal{S}(\mathcal{C}_q,{\underline \phi}) \subset\mathbb{R}^{\ell n}$ is a random codebook of size $q^k$. Thus the RS code $\mathcal{C}_q$ can be mapped to $(q!)^n$ different codebooks $\mathcal{S}$, each of which with probability ${\rm Pr}\{\mathcal{S}\} = 1/(q!)^n$.

Given a codebook $\mathcal{S}(\mathcal{C}_q, {\underline \phi})$, we denote
$B_{\delta}(\mathcal{S})$ the average number of {\em ordered} pairs
of codewords with Euclidean distance $\delta$ in
$\mathcal{S}(\mathcal{C}_q,{\underline \phi})$. We define
\begin{equation}\label{Bd}
B_{\delta} \stackrel{\Delta}{=} \sum_{\mathcal{S}}{\rm
Pr}\{\mathcal{S}\}B_{\delta}(\mathcal{S}),
\end{equation}
which denotes the ensemble average of the number of ordered pairs of
codewords with Euclidean distance $\delta$.
\begin{definition}
The {\em average Euclidean distance enumerating function} of RS-CM
with random modulation is defined as
\begin{equation}\label{B(X)}
B(X) \stackrel{\Delta}{=} \sum_{\delta}B_{\delta}X^{\delta^2}.
\end{equation}
We call $\{B_{\delta}\}$ the average Euclidean distance spectrum.
\end{definition}

\subsection{Analytic Bounds for the Ensemble of RS-CM with Random Modulation}
The ML decoding error probability ${\rm Pr}\{E\}$ for the ensemble
of RS-CM can be written as
\begin{equation}\label{Pr(E)}
{\rm Pr} \{E\}=\sum_{\mathcal{S}}{\rm Pr}\{\mathcal{S}\}{\rm
Pr}\{E|\mathcal{S}\},
\end{equation}
where ${\rm Pr}\{E|\mathcal{S}\}$ is the conditional ML decoding
error probability given a code $\mathcal{S}(\mathcal{C}_q, {\underline \phi})$.

From~(\ref{UB}), the UB of ${\rm Pr}\{E|\mathcal{S}\}$ is
\begin{equation}
{\rm Pr} \{E|\mathcal{S}\}\leq
\sum_{\delta}B_{\delta}(\mathcal{S})Q\left(\frac{\delta}{2\sigma}\right).
\end{equation}
Therefore, from~(\ref{Bd}) and~(\ref{Pr(E)}), the UB on the ML
decoding error probability of the ensemble of RS-CM can be written as
\begin{eqnarray}\label{AUB}
{\rm Pr} \{E\}&=&\sum_{\mathcal{S}}{\rm Pr}\{\mathcal{S}\}{\rm
Pr}\{E|\mathcal{S}\} \nonumber\\
&\leq&\sum_{\mathcal{S}}{\rm
Pr}\{\mathcal{S}\}\sum_{\delta}B_{\delta}(\mathcal{S})Q\left(\frac{\delta}{2\sigma}\right) \nonumber\\
&=&\sum_{\delta}\sum_{\mathcal{S}}{\rm
Pr}\{\mathcal{S}\}B_{\delta}(\mathcal{S})Q\left(\frac{\delta}{2\sigma}\right) \nonumber\\
&=&\sum_{\delta}B_{\delta}Q\left(\frac{\delta}{2\sigma}\right),
\end{eqnarray}
which is determined by the average Euclidean distance spectrum
$\{B_{\delta}\}$.

From~(\ref{SB}), the SB of ${\rm Pr}\{E|\mathcal{S}\}$ is
\begin{equation}
{\rm Pr} \{E|\mathcal{S}\}\leq
\int_{0}^{+\infty}\min\left\{f_u(r|\mathcal{S}),1\right\}g(r)~{\rm
d}r,
\end{equation}
where $g(r)$ is given by~(\ref{gr}) and $f_u(r|\mathcal{S})=\sum_{\delta}B_{\delta}(\mathcal{S}) p_2(r,
    \delta)$ with $p_2(r, \delta)$ given by~(\ref{p2}). From~(\ref{Bd}), we define
\begin{eqnarray}\label{fu(r)}
f_u(r)&\triangleq&\sum_{\mathcal{S}}{\rm Pr}\{\mathcal{S}\}f_u(r|\mathcal{S}) \nonumber\\
&=&\sum_{\mathcal{S}}{\rm Pr}\{\mathcal{S}\}\sum_{\delta}
B_{\delta}(\mathcal{S}) p_2(r,
    \delta) \nonumber\\
&=&\sum_{\delta}\sum_{\mathcal{S}}{\rm
Pr}\{\mathcal{S}\}B_{\delta}(\mathcal{S}) p_2(r,
    \delta) \nonumber\\
&=&\sum_{\delta}B_{\delta} p_2(r,
    \delta).
\end{eqnarray}
Therefore, from~(\ref{Pr(E)}) and~(\ref{fu(r)}), the SB on the ML
decoding error probability of the ensemble of RS-CM can be written as
\begin{eqnarray}\label{ASB}
{\rm Pr} \{E\}&=&\sum_{\mathcal{S}}{\rm Pr}\{\mathcal{S}\}{\rm
Pr}\{E|\mathcal{S}\} \nonumber\\
&\leq&\sum_{\mathcal{S}}{\rm
Pr}\{\mathcal{S}\}\int_{0}^{+\infty}\min\left\{f_u(r|\mathcal{S}),1\right\}g(r)~{\rm
d}r \nonumber\\
&\leq&\int_{0}^{+\infty}\min\left\{\sum_{\mathcal{S}}{\rm
Pr}\{\mathcal{S}\}f_u(r|\mathcal{S}),1\right\}g(r)~{\rm
d}r \nonumber\\
&\leq&\int_{0}^{+\infty}\min\left\{f_u(r),1\right\}g(r)~{\rm d}r,
\end{eqnarray}
which is determined by the average Euclidean distance spectrum
$\{B_{\delta}\}$.

\subsection{Computation of the Average Euclidean Distance Enumerating Function}

As seen from the preceding subsection, computing the derived bounds for
the ensemble of RS-CM requires the average Euclidean distance enumerating function $B(X)$ defined in~(\ref{B(X)}). In this subsection, we will show that $B(X)$ is computable from the Hamming weight enumerating function $W(X)$ and the Euclidean distance enumerating function of the signal constellation.

Recall that $\mathcal{X}$ is the signal constellation of size $q$. Define
\begin{equation}\label{D(X)}
  D(X) \stackrel{\Delta}{=} \sum_{\delta}D_{\delta} X^{\delta^2} = \frac{1}{q(q-1)} \sum_{x \in \mathcal{X}, y \in \mathcal{X}, x\neq y} X^{||x-y||^2},
\end{equation}
where $||\cdot||$ denotes the Euclidian distance and $D_{\delta}$ denotes the average number of {\em ordered} signal
pairs with Euclidean distance $\delta$ in the constellation
$\mathcal{X}$. Let $\underline c$ and $\hat{\underline c}$ be two codewords with Hamming distance $d > 0$. Denote by ${\underline \phi} = (\phi_0, \phi_1, \cdots, \phi_{n-1})$ the sequence of random mappings, whose components are independent and uniformly distributed over $\Phi$, the set of all one-to-one mappings. Then we have,
 \begin{equation*}
    \sum_{\underline \phi}(q!)^{-n} X^{\sum_t ||\phi_t(c_t) - \phi_t({\hat c}_t)||^2} = (D(X))^d.
 \end{equation*}
Therefore,
\begin{eqnarray*}
  B(X) &=& \sum_{\underline \phi}(q!)^{-n}\sum_{\underline c \in \mathcal{C}_q}q^{-k} \sum_{\hat{\underline c} \in \mathcal{C}_q, \underline c \neq \hat{\underline c}} X^{\sum_t||\phi_t(c_t) - \phi_t(\hat{c}_t)||^2}\\
  &=& \sum_{\underline c \in \mathcal{C}_q}q^{-k} \sum_{\hat{\underline c} \in \mathcal{C}_q, \underline c \neq \hat{\underline c}} \sum_{\underline \phi}(q!)^{-n}X^{\sum_t||\phi_t(c_t) - \phi_t(\hat{c}_t)||^2}\\
   &=& \sum_{d_{\min} \leq d \leq n} W_d \cdot (D(X))^d.
\end{eqnarray*}

{\bf Remark.} When considering RS-CM using BPSK modulation, the
average Euclidean distance enumerating function $B(X)$ is reduced
to the average binary weight enumerator~(BWE) presented
in~\cite{Mostafa04}.

\begin{figure}[!t]
\centering
\includegraphics[width = 9cm]{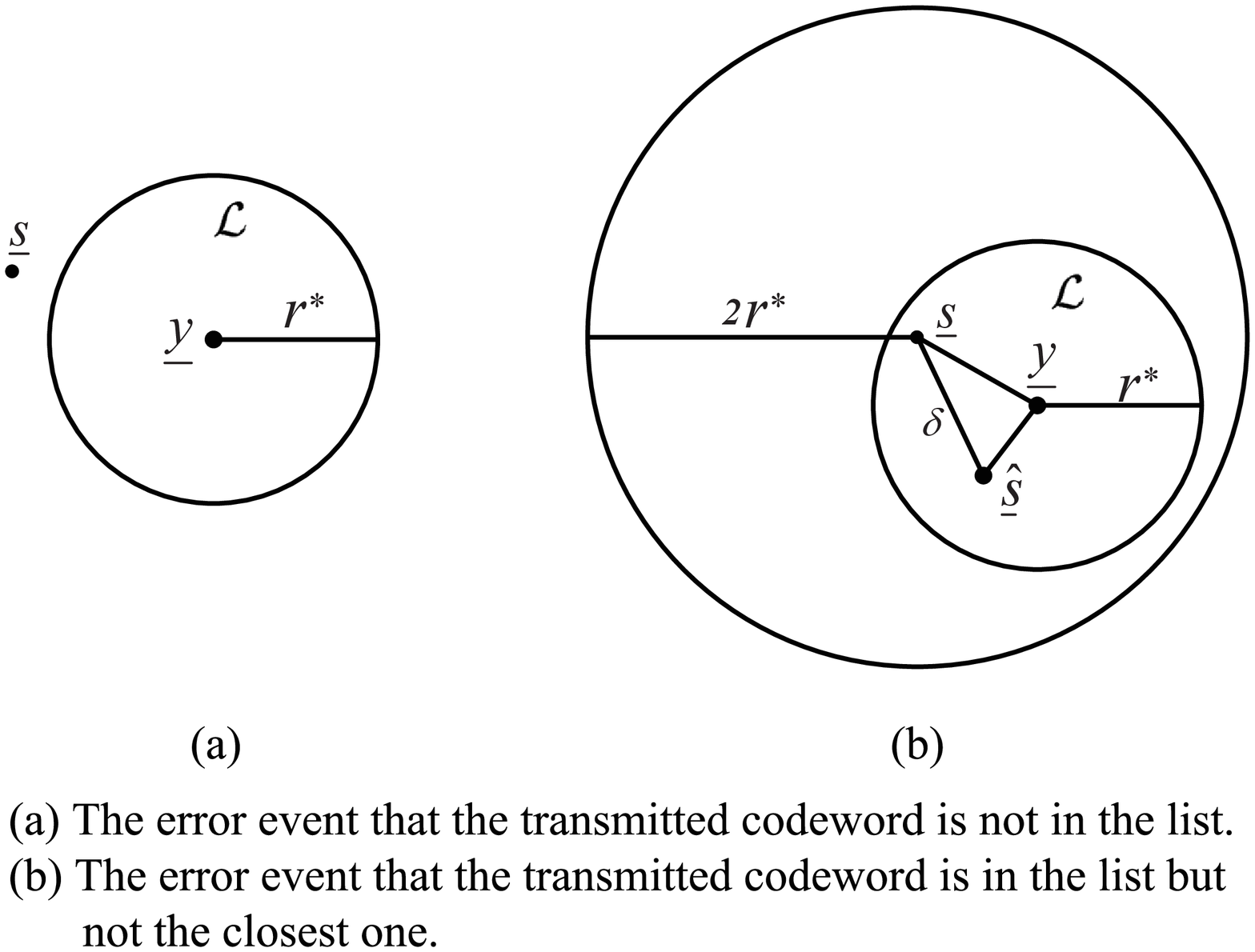}
\caption{Graphical illustrations of the decoding error events.}
\label{Error}
\end{figure}

\section{Simulation-based Bounds for Specific RS-coded Modulation System}\label{sec4}

In this section, we present a simulation-based bound for specific RS-CM system by the aid of a list decoding algorithm.

\begin{algorithm}{A suboptimal list decoding algorithm for the purpose of
performance analysis}\label{subDEC}
\begin{enumerate}
  \item[S1.] List all codewords within the Euclidean sphere with center
at the received vector $\underline{y}$ and with radius $r^{*}\geq 0$
where $r^*$ is a parameter to be determined. We denote the list as
$\mathcal{L}$.
  \item[S2.] Find the codeword $\underline{s}^*\in
\mathcal{L}$ that is closest to $\underline{y}$.
\end{enumerate}
\end{algorithm}

{\bf Remark.} The above list decoding algorithm was also referred to as sphere decoding algorithm in~\cite{El-Khamy09}. The objective of~\cite{El-Khamy09} is to derive analytic bounds on the sphere decoding itself, while our objective here
is to derive simulation-based bounds on the ML decoding by assuming that the sphere decoding can be implemented efficiently.

The decoding error occurs in two cases under the assumption that the
codeword ${\underline s}$ is transmitted.

{\em Case 1}. The transmitted codeword ${\underline s}$ is not in
the list $\mathcal{L}$~(see Fig.~\ref{Error}~(a)), that is,
$\|\underline{z}\|=\|\underline{y}-\underline{s}\| \geq r^*$. This event
is denoted by $\{E_1|\underline{s}\}$, whose probability is given by
\begin{eqnarray}\label{B2}
{\rm Pr} \{E_1|\underline{s}\} &=& {\rm
Pr}\{\|\underline{z}\| \geq r^*\} \nonumber\\
&=& \int_{r^*}^{+\infty}g(r)~{\rm d}r,
\end{eqnarray}
where $g(r)$ is defined in~(\ref{gr}).

{\em Case 2}. The transmitted codeword ${\underline s}$ is in the
list $\mathcal{L}$, but is not the closest one~(see
Fig.~\ref{Error}~(b)). The event is denoted by
$\{E_2|\underline{s}\}$, resulting in the error probability ${\rm
Pr}\{E_2|\underline{s}\}$.


Obviously, ${\rm Pr}\{E_2|\underline s\} \leq {\rm Pr}\{E|\underline s\} \leq {\rm Pr}\{E_1|\underline s\} + {\rm Pr}\{E_2|\underline s\}$. Averaging over the transmitted codewords, we have
\begin{equation}\label{SimB}
{\rm Pr} \{E_2\}\leq {\rm Pr} \{E\} \leq {\rm Pr} \{E_1\}+{\rm Pr}\{E_2\},
\end{equation}
where ${\rm Pr} \{E_1\} = \sum_{\underline{s}}{\rm Pr}\{\underline{s}\}{\rm
Pr} \{E_1|\underline{s}\} = \int_{r^*}^{+\infty}g(r)~{\rm d}r$ and $
{\rm Pr} \{E_2\}=\sum_{\underline{s}}{\rm Pr}\{\underline{s}\}{\rm
Pr} \{E_2|\underline{s}\}$. If we can calculate ${\rm Pr}\{E_2\}$, then we have bounds on the ML decoding error probability. Actually, for small $r^*$ or short RS codes, this probability ${\rm Pr}\{E_2\}$ can be estimated by {\em Monte Carlo} simulation using the recently proposed tree-based Chase-type algorithm~\cite{Tang13x}.

\begin{algorithm}
Estimate the error probability ${\rm Pr} \{E_2\}$.
\begin{algorithmic}[1]
\STATE Initialize $i=0$ and $N_{err} = 0$. Given a parameter $r^* > 0$ and a sufficiently large integer $N_{total} > 0$.
\WHILE {$(i<N_{total})$}
    \STATE Generate uniformly at random a codeword $\underline{s}$ and a white Gaussian noise sample $\underline{z}$.
    \STATE $\underline{y}\leftarrow \underline{s}+\underline{z}$
    \IF {$\|\underline{y}-\underline{s}\|\leq r^*$}
        \STATE $i\leftarrow i+1$
        \STATE Decode $\underline y$ with the algorithm~\cite{Tang13x}, resulting in the decoded codeword $\underline{s}^*$.
        \IF {$\underline{s}^*$ is different from $\underline{s}$}
            \STATE $N_{err}\leftarrow N_{err} + 1$
        \ENDIF
    \ENDIF
\ENDWHILE
\STATE ${\rm Pr} \{E_2\} = N_{err} / N_{total}$
\RETURN ${\rm Pr} \{E_2\}$
\end{algorithmic}
\end{algorithm}

{\bf Remark.} Notice that the above algorithm can be faster than the real decoding algorithm since we do not have to find the optimal candidate codeword within the sphere in the case that some {\em earlier} intermediate candidate is found to be better than the transmitted one\footnote{In this case, the ML decoding must also make an error, as pointed out in~\cite{Dorsch74} and used in~\cite{Valembois04}.}.

\begin{figure}
\centering
  \includegraphics[width=12cm]{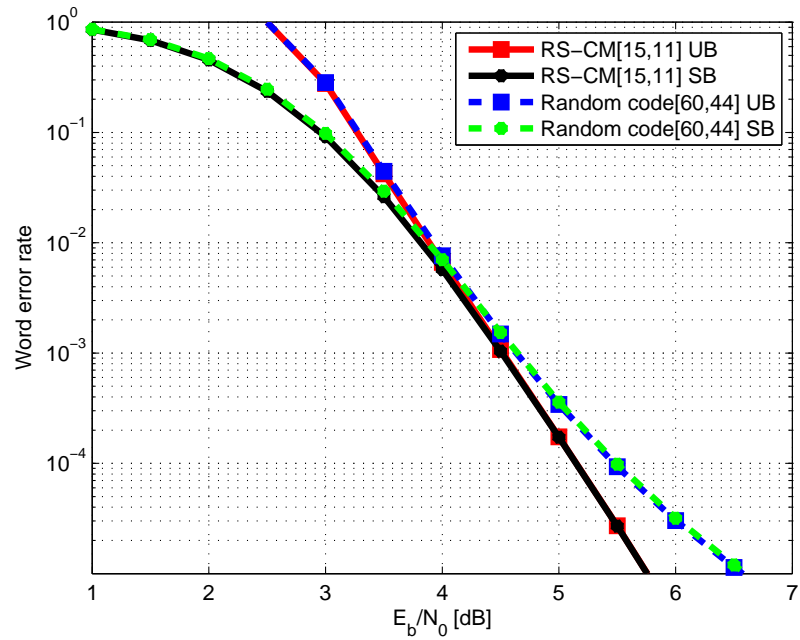}\\
  \caption{Upper bounds on the word error rate~(WER) of the ensemble of $\mathcal{C}_{16}[15,11,5]$ RS-CM using BPSK modulation.}\label{Bound1}
  \end{figure}


\begin{figure}
\centering
  \includegraphics[width=12cm]{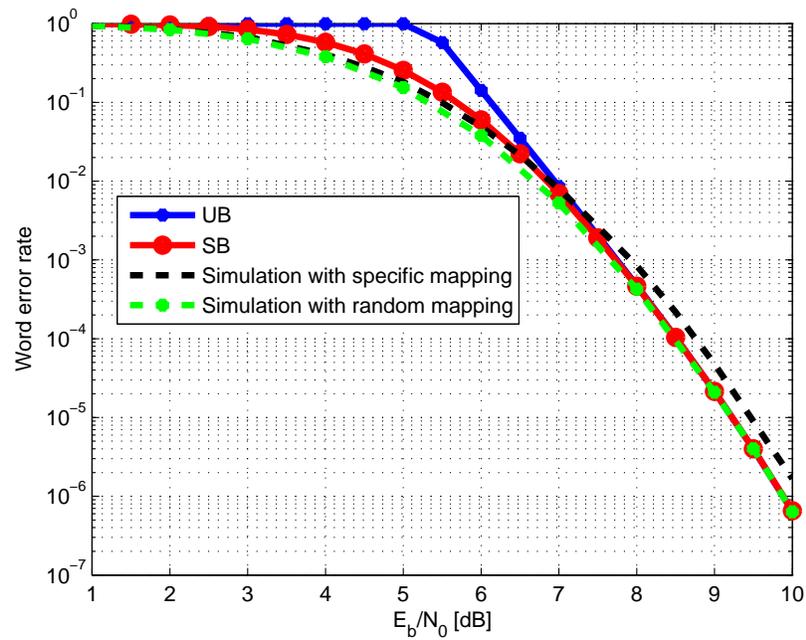}\\
  \caption{Upper bounds on the WER of the ensemble of $\mathcal{C}_{16}[15,11,5]$ RS-CM using $16$-QAM signal constellation.}\label{Bound2}
  \end{figure}

\begin{figure}
\centering
  \includegraphics[width=12cm]{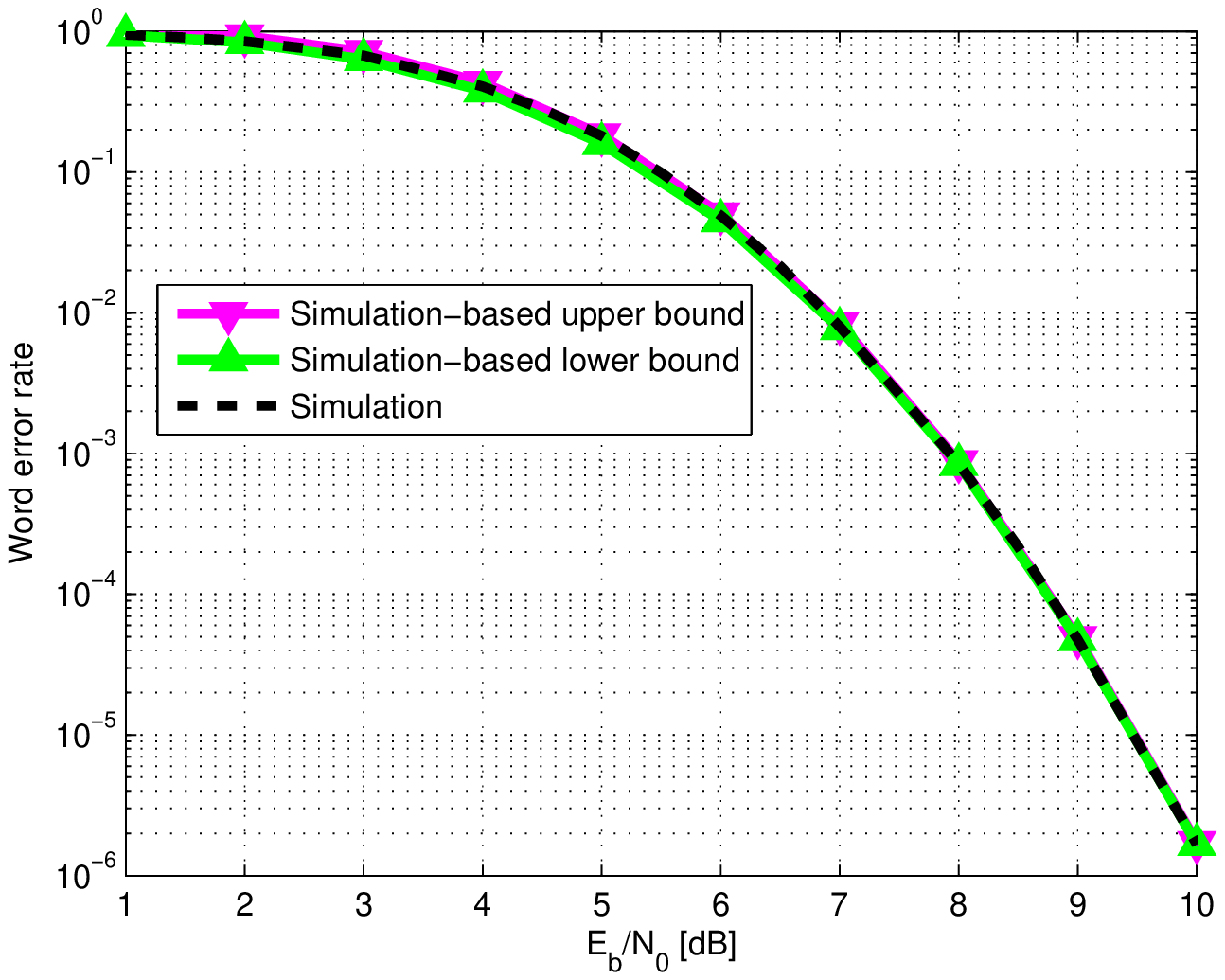}\\
  \caption{Simulation-based bounds on the WER of $\mathcal{C}_{16}[15,11,5]$ RS-CM using $16$-QAM signal constellation.}\label{Bound3}
  \end{figure}

\section{Numerical Results}\label{sec5}

{\em Example 1}: In this example, we consider $\mathcal{C}_{16}[15, 11, 5]$ using BPSK modulation. For comparison, we also consider a random binary linear code $\mathcal{C}_2[60, 44]$ with BPSK modulation. Both these two codes have the same code length and code rate. The numerical results are shown in Fig.~\ref{Bound1}. We can see that the SB is tighter than the UB in the low signal-to-noise ratio~(SNR) regime, as expected. The results also show that this nonbinary structured code is better~(on average) than binary random linear codes. Since binary linear codes can achieve the capacity as the code length goes to infinity, this numerical result indicates that, in short-length regime, the code structure plays a more important role.

{\em Example 2}: In this example, we consider the same code as in Example~1 but using $16$-QAM modulation. The numerical results are shown in Fig.~\ref{Bound2}. We can see that the SB is tighter than the UB in the low SNR regime. We can also see that, in the high SNR regime, we do not have to use the complicated SB to predict the performance. Also presented in Fig.~\ref{Bound2} are the simulation results with specific/random modulation using the recently proposed tree-based Chase-type algorithm~\cite{Tang13x}. It can be seen that the simulated curve with random mapping matches well with the analytic bounds for the ensemble and the simulated curve with specific mapping is slightly worse than the ensemble upper bounds in the high SNR regime. This comparison also shows that the performance of RS-CM is closely
related to which modulation is chosen.

{\em Example 3}: In this example, we consider the same RS-CM system as in Example~2 but using specific modulation. The numerical results are shown in Fig.~\ref{Bound3}. Also presented in Fig.~\ref{Bound3} are the simulation results using the same decoding algorithm~\cite{Tang13x} as in Example~2. It can be seen that all these curves are almost overlapped, showing that the decoding algorithm is near optimal.

\section{Conclusions}\label{conclusion}

In this paper, we have presented analytic bounds for the ensemble of
RS-CM systems using random modulation. We have also presented simulation-based bounds for any specific RS-CM systems. Numerical results show that, at least in the short code-length regime, the performance of RS-CM with random modulation is better than that of random linear codes. Numerical results also show that the recently proposed Chase-type decoding algorithm is near optimal.

%
%
%

\section*{Acknowledgment}

The authors would like to thank Siyun Tang for helpful discussions in connection to the simulation-based bounds.


\small
\bibliographystyle{IEEEtran}
\end{document}